\documentclass{amia}
\usepackage{graphicx}
\usepackage[labelfont=bf]{caption}
\usepackage[superscript,nomove]{cite}
\usepackage{color}
\usepackage{todonotes}
\usepackage{boldline, makecell}
\usepackage{amsmath, amssymb, url, bbm}

\usepackage[symbol]{footmisc}

\begin{document}

\title{Weak Supervision for Affordable Modeling of Electrocardiogram Data}

\author{Mononito Goswami, B.Tech.$^{1}$, Benedikt Boecking, M.Sc.$^{1}$, Artur Dubrawski, Ph.D.$^{1}$}

\institutes{
    $^1$Auton Lab, School of Computer Science, Carnegie Mellon University\\ Pittsburgh, PA, USA\\
}

\maketitle

\noindent{\bf Abstract}

\textit{Analysing electrocardiograms (ECGs) is an inexpensive and non-invasive, yet powerful way to diagnose heart disease. ECG studies using Machine Learning to automatically detect abnormal heartbeats so far depend on large, manually annotated datasets. While collecting vast amounts of unlabeled data can be straightforward, the point-by-point annotation of abnormal heartbeats is tedious and expensive. We explore the use of multiple weak supervision sources to learn diagnostic models of abnormal heartbeats via human designed heuristics, without using ground truth labels on individual data points. Our work is among the first to define weak supervision sources directly on time series data. Results show that with as few as six intuitive time series heuristics, we are able to infer high quality probabilistic label estimates for over 100,000 heartbeats with little human effort, and use the estimated labels to train competitive classifiers evaluated on held out test data.}

\section*{Introduction} 
\label{sec:intro}
% Cardiovascular diseases accounted for almost $31\%$ global deaths in $2016$ alone. 
Automatic analysis of electrocardiograms (ECGs) 
promises substantial improvements in critical care. ECGs offer an inexpensive and non-invasive way to diagnose irregularities in heart functioning. \textit{Arrhythmias} are abnormal heartbeats which alter both the morphology and frequency of ECG waves, and can be detected in an ECG exam. However, identifying and classifying arrhythmias manually is not only error-prone but also cumbersome. Clinicians may have to analyze each heartbeat in an ECG record, and in critical care settings, carefully analysing each heartbeat is nearly impossible. As a consequence, the medical machine learning (ML) community has worked extensively on computational models to automatically detect and characterize arrhythmias~\cite{ebrahimi2020review, luz2016ecg}.

Rajpurkar et al. \cite{hannun2019cardiologist} demonstrated that modern ML models trained on a large and diverse corpus of patients can exceed the performance of certified cardiologists in detecting abnormal heartbeats. But their Convolutional Neural Network model was trained on a manually annotated dataset of more than 64,000 ECG records from over 29,000 patients. Clearly, research on automated arrhythmia detection has moved the burden of monitoring ECG in critical care to annotating and curating large databases on which ML models can be trained and validated. 
This currently prevailing process involves laborious manual data labeling that is a major bottleneck of supervised medical ML applications in practice. 
Popular ML techniques, in particular deep learning, require a large supply of reliably annotated training data, containing records from a diverse cohort of patients.
According to Moody et al.~\cite{moody2001impact} and our own experience, %-- who prepared the MIT-BIH Arrhythmia Database--% over a period of $5$ years 
raw medical data is abundant, but its thorough characterization can be involved and expensive. 
This reliance on labeled data forces researchers to often use static and older datasets, despite evolving patient populations, systematic improvements in understanding of diseases, and advances in medical equipment. 

%While advances in technology since the late 1970s have been substantial, 
Recent developments in e.g.\ web-based tools to visualize and annotate ECG signals have not reduced the annotation time and effort significantly. 
For example, it took $4$ doctors, almost $3$ months to annotate $15,000$ short ECG records using the LabelECG tool~\cite{ding2019labelecg}. 
In general, gold standard expert annotations can be costly. Conservative estimates place the hourly cost of highly qualified labor for the related task of EEG annotation between $\$50$ and $\$200$ per hour~\cite{abend2015much}. 

% , despite them not having to segment data into beats and waves by hand

\begin{figure*}[!htbp]
    \centering
    \includegraphics[width = .9\textwidth]{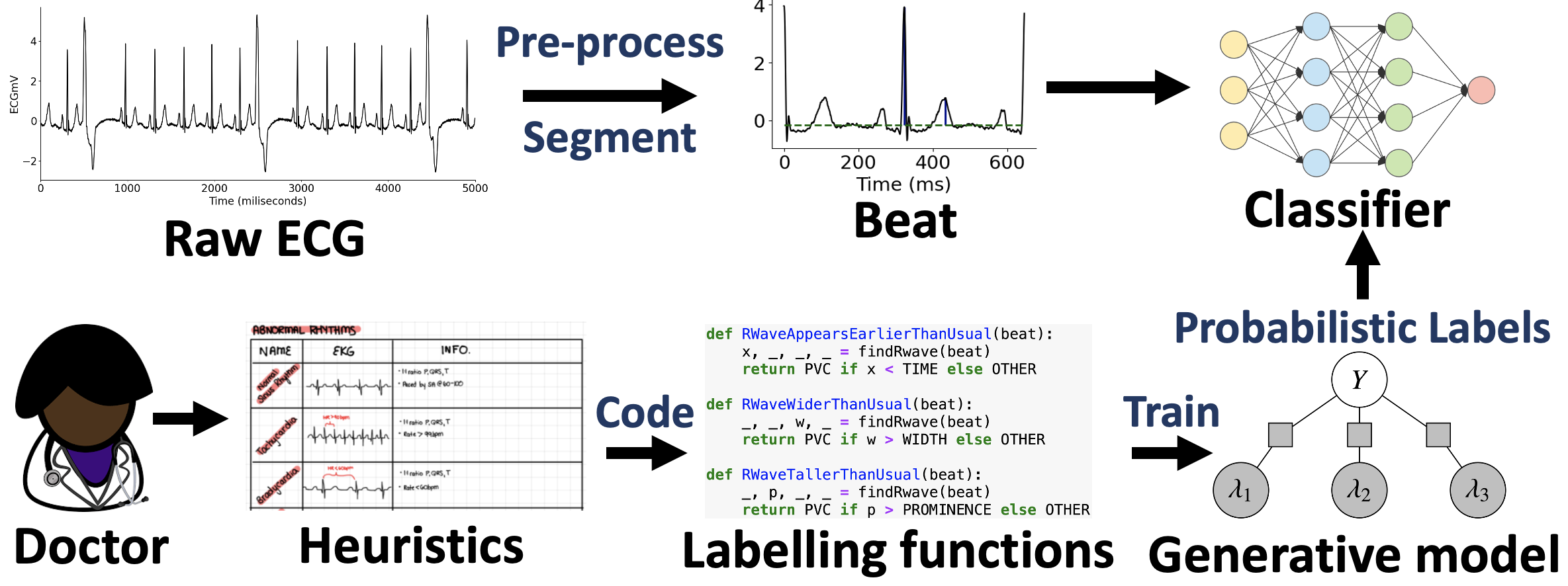}
    \caption{Data programming with time series heuristics can affordably train competitive end models for automated ECG adjudication. Instead of labeling each data point by hand (fully supervised setting), experts encode their domain knowledge using noisy labeling functions (LFs). A label model then learns the unobserved empirical accuracy of LFs and uses them to produce probabilistic data label estimates using weighted majority vote.}
    \label{fig:workflow}
\end{figure*}

In this work, we explore the use of multiple cheaper albeit perhaps noisier supervision sources to learn an arrhythmia detector, without access to ground truth labels of individual samples. 
We follow the recently proposed \textit{data programming (DP)}~\cite{ratner2016data} framework in which a factor graph is used to model user-designed heuristics to obtain a probabilistic label for each heartbeat instance. 
DP has gained attention from the medical imaging and general ML community and has been used for various tasks such as automated detection of seizures from electroencephalography \cite{saab2020weak}, intracranial hemorrhage detection with computed tomography, or automated triage of Extremity Radiograph Series \cite{dunnmon2020cross}. 

Our experiments with ECG data from the MIT-BIH Arrhythmia Database indicate that with as few as $6$ heuristics, we are able to train an arrhythmia detection model with only a small amount of human effort.
The resulting model is competitively accurate when compared to a model trained on the same data with full supply of pointillistic ground truth annotations. 
It can also outperform another alternative  model trained using active learning, a popular technique used to reduce data labeling efforts when they are expensive. 
We also show that domain heuristics can be automatically tuned to account for inter-patient variability and further boost reliability of the resulting models.

While many different types of arrhythmias exist, for illustration purposes we focus on identifying heartbeats showing Premature Ventricular Contractions (PVCs). Whereas isolated infrequent PVCs are usually benign, frequent PVCs with exceptionally wide QRS complexes\footnote[2]{QRS complexes are generally the most prominent spike seen on a typical ECG. They are a combination of the Q wave, R wave and S wave, which occur in rapid succession, and represent an electrical impulse.} may be indicative of heart disease and eventually lead to sudden cardiac death~\cite{moulton1990premature}. 
However, our approach is general and applicable to all classes of abnormal heartbeats. 

%The rest of the paper is organized as follows. In the next section, we provide a brief overview of related work. Later in the Methodology section we describe the creation of domain heuristics and the modeling approach. We follow it with the Experiments and Results section describing the data, experimental setup, and our findings. Finally, we conclude our paper with a brief discussion and avenues of future work. 

\section*{Related Work}
\paragraph{Automated Arrhythmia Detection} Automatically detecting abnormal heartbeats is a widely studied problem. Most researchers in the past relied on manually labeled corpora such as the MIT-BIH Arrhythmia Database, the AHA Database for Evaluation of Ventricular Detectors, etc., to train and validate their models~\cite{ebrahimi2020review, luz2016ecg}. Rajpurkar et al.~\cite{hannun2019cardiologist} recently demonstrated that a deep Convolutional Neural Network (CNN) can even exceed the performance of experienced cardiologists. However, their model was trained on as many as 64,121 thirty-second ECG records from 29,163 patients, manually-labeled by a group of certified cardiographic technicians. 
Hence, to fuel advances in automated arrhythmia detection and, more generally, in ML-aided healthcare, there is a clear need to affordably label vast amounts of data. 

Some recent studies have attempted to address the annotation bottleneck, albeit at a different context and scale. These studies have used semi-supervised or active learning to incrementally improve the accuracy of models without significant expert intervention. For instance, to overcome inter-patient variability without additional manual labeling of patient specific data, Zhai et al.~\cite{zhai2020semi} iteratively updated the preliminary predictions of their trained CNN using a semi-supervised approach. Correspondingly, Wang et al. \cite{wang2019global} used active learning on newly acquired data to choose the most informative unlabeled data points and incorporate them in the training set. Sayantan et al.~\cite{sayantan2018classification} used active learning to improve their model's classification results with the help of an expert. So far, the work which comes closest to addressing the problem of intelligently labeling vast quantities of ECG data is that of Pasolli et al.~\cite{pasolli2010active}. Starting from a small sub-optimal training set, the authors proposed three active learning strategies to choose additional heartbeat instances to further train an Support Vector Machine (SVM) model. Their work demonstrated that models trained using active learning can achieve impressive performance while using few labeled samples. In this work, we also compare the performance of our weakly supervised method with active learning. As against Pasolli et al.~\cite{pasolli2010active} who used a manually curated set of ECG features and trained an SVM using margin sampling, we train a Convolutional Neural Network (CNN) which can automatically learn rich feature representations using uncertainty sampling, another popular active learning strategy. 

\paragraph{Weak Supervision} Of late, some studies have explored the use of multiple noisy heuristics to programmatically label data at scale. The recently proposed Data Programming framework~\cite{ratner2016data}, where experts express their domain knowledge in terms of intuitive yet perhaps noisy labeling functions (LFs), is a prominent example. Recent studies have used DP for a wide range of clinical applications, ranging from detecting aortic valve malformations using cardiac MRI sequences \cite{fries2019weakly}, seizures using EEG \cite{saab2020weak} and brain hemorrhage using 3D head CT scans \cite{dunnmon2020cross}. With the exception of Khattar et al.~\cite{khattar2019multi}, most prior work on DP has been on image~\cite{dunnmon2020cross, fries2019weakly} or natural language~\cite{ratner2017snorkel} modalities. Moreover, prior DP research either used weak annotations from lab technicians, students~\cite{saab2020weak}, or heuristics built on auxiliary modalities (e.g., clinician notes, text reports \cite{dunnmon2020cross}, patient videos \cite{saab2020weak}), some of which only allow for coarse annotation of the entire time series rather than of the individual segments. On the contrary, we define heuristics directly on time series. This enables seamless labeling of entire time series or their segments using the same framework.   

\section*{Methodology}\label{sec:methods}
In this section, we will describe how we use domain knowledge to define heuristics to detect PVC in ECG time series. These heuristics will noisily label subsets of data. We will model these noisy labels to obtain an estimate of the unknown true class label for each data point. We then use the estimated labels to train the final classifier, which will be evaluated on held out test data and compared to alternative models trained using ground-truth labels directly. Fig.~\ref{fig:workflow} describes the full workflow we follow to train the end model $f$. 
%We will first briefly describe the clinical background on PVC detection and then explain the modeling process and heuristic definitions in detail. 

% \begin{figure*}[!htbp]
%     \centering
%     \includegraphics[width=0.65\textwidth]{PVCandNormalbeat.png}
%     \caption{Examples of a PVC (left) and normal (right) heartbeat. The dotted, green horizontal lines represent the detected ECG baselines, while the blue and red vertical lines mark the QRS-complexes and T-waves, respectively.}
%     \label{fig:PVCandNormalbeat}
% \end{figure*}

\begin{figure*}[!hbtp]
    \centering
  \begin{tabular}{cc}
    \includegraphics[width=0.25\textwidth]{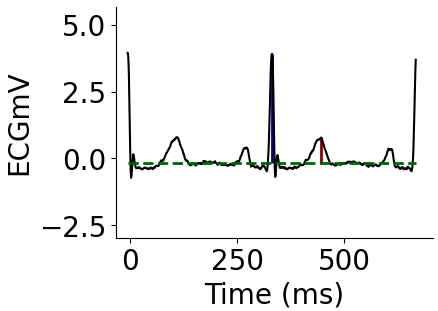} 
    &\hspace{15mm}\includegraphics[width=0.23\textwidth]{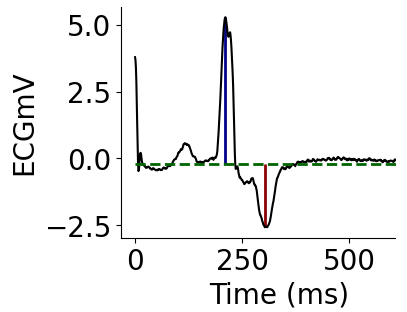} \\
    (i)&\hspace{10mm}(ii) \\
  \end{tabular}
    \caption{Examples of a normal (i) and PVC (ii) heartbeat. Dotted green horizontal lines represent the ECG baselines detected during pre-processing, blue and red vertical lines mark the QRS-complexes and T-waves.}
    \label{fig:PVCandNormalbeat}
\end{figure*}

\noindent\paragraph{Domain Knowledge to Identify PVC}
\label{sec:heuristics}
A Premature Ventricular Contraction is a fairly common event when the heartbeat is initiated by an impulse from an ectopic focus which may be located anywhere in the ventricles rather than the sinoatrial node. On the ECG, a PVC beat appears earlier than usual with an abnormally tall and wide QRS-complex, with its ST-T vector directed opposite to the QRS vector, Fig.~\ref{fig:PVCandNormalbeat}(ii). 
These generic characteristics allowed one non-domain-expert user to define $6$ heuristics in less than $30$ minutes. The user was initially unfamiliar with clinical ECG interpretation and referred to an online textbook \cite{PVCreference} to develop heuristics.
% \url{https://ecgwaves.com/course/the-ecg-book/} 
Expert clinicians are likely able to define heuristics more rapidly and thoroughly. 
%Each heuristic serves as a labeling function for the DP framework, described in the next section. 
The heuristics listed below are defined directly on time series. This is in contrast to prior work which uses weak annotations or heuristics defined on an auxiliary modality such as text or images. 

\textbf{Heuristics:} i.\ R-wave appears earlier than usual, ii.\ R-wave is taller than usual, iii.\ R-wave is wider than usual, iv.\ QRS-vector is directed opposite to the ST-vector, v.\ QRS-complex is inverted, vi.\ Inverted R-wave is taller than usual. 

% \begin{table*}[!htbp]
%     \centering
%     \resizebox{0.9\textwidth}{!}{
%     \begin{tabular}{c c c c c}
%          \Xhline{1pt}
%          \textbf{labeling Function} & \textbf{Emp. Acc. (\%)} & \textbf{Conflicts (\%)} & \textbf{Sens. drop} & \textbf{Spec. drop}\\ \hline
%          \texttt{RWaveAppearsEarlierThanUsual} & 93.96 & 72.74 & -8.13 & -6.19 \\
%          \texttt{RWaveWiderThanUsual} & 84.90 & 72.74 & -75.4 & 1.66 \\
%          \texttt{RWaveTallerThanUsual} & 93.96 & 72.74 & -10.83 & -4.89 \\
%          \texttt{QRSComplexSTTVectorOpposite} & 37.57 & 71.30 & -41.61 & 0.19\\
%          \texttt{InvertedQRSComplex} & 22.06 & 9.88 & -41.33 & -0.62 \\
%          \texttt{InvertedRWaveWiderThanUsual} & 48.46 & 19.13 & -45.55 & -1.44 \\ \Xhline{1pt}
%     \end{tabular}}
%     \caption{labeling function (LF) evaluation results. Coverage is the percentage of the dataset where the LF does not abstain, whereas empirical accuracy is the accuracy with respect to the ground truth labels.} 
%     \label{tab:LFsresults}
% \end{table*}
\paragraph{Modeling Labeling Functions over Patient Time Series}
We will now describe our formal assumptions about the dataset and heuristics, and introduce the modeling procedure. Given an ECG dataset of $p$ patients $X=\{x^j\}_{j=1}^p$, where $x^j \in \mathbb{R}^{T}$ are raw ECG vectors of length $T$, we can segment each ECG $x^j$ into $B<T$ beats such that $x^j=\{x^j_1,\dots,x^j_B\}$. Each segment $b \in \{1,\dots,B\}$ has an unknown class label $y_b \in \{-1,1\}$, where $y^j_b=1$ represents a premature ventricular contraction (\texttt{PVC}).  
Our goal is to use domain knowledge to model the unknown $y^j_b$, without having to annotate the instances individually, to then train an end classifier $f(x^j_b)=y^j_b$ for automatic detection of PVC. 
We define $m$ labeling functions (LFs) $\{\lambda_h(x^j_b)\}_{h=1}^{m}$ directly on the time series. These LFs noisily label subsets of beats with $\lambda_h(x^j_b)=\{-1,0,1\}$ corresponding to votes for \textit{negative, abstain, or positive}. These LFs do not have to be perfect and may conflict on some samples, but must have accuracy better than random \cite{boecking2020interactive}. DP uses this voting behavior to infer true labels by learning the empirical accuracies, propensities and, optionally, dependencies of the LFs via a factor graph. 
We use a factor graph as introduced in Ratner et al.~\cite{ratner2016data} to model the $m$ user defined labeling functions. For simplicity, we assume that the LFs are independent conditioned on the unobserved class label.
Let $Y^j=(y^j_1,\dots,y^j_B) \in \{-1,1\}^B$ be the concatenated vector of the unobserved class variable for the $B$ beat segments of patient $j$ and $\Lambda^j=\{-1,0,1\}^{B \times m}$ be the LF output matrix where $\Lambda^j_{ik}=\lambda_k(x^j_i)$ is the output of LF $k$ on beat $i$ of patient $j$. 
We define a factor for LF accuracy as 
\begin{equation*}
\phi_{i,k}^{Acc}(\Lambda,Y) \triangleq \mathbbm{1}\{\Lambda_{ik}=y_i\}
\end{equation*}
We also define a factor of LF propensity as
\begin{equation*}
\phi_{i,k}^{Lab}(\Lambda,Y) \triangleq \mathbbm{1}\{\Lambda_{ik}\neq 0\}
\end{equation*}
Then, the label model for a patient $j$ is defined as
\begin{align}
\begin{split}\label{eq:labelmodel}
 p_{\theta}( Y^j, \Lambda^j  )
&\triangleq Z_{\theta}^{-1}
\exp (\sum_{i=1}^B \sum_{k=1}^m\theta_k \phi_{i,k}^{Acc}(\Lambda^j_i,y^j_i) + \sum_{i=1}^B \sum_{k=1}^m\theta_k \phi_{i,k}^{Lab}(\Lambda^j_i,y^j_i))
\end{split}
\end{align}
where $Z_{\theta}$ is a normalizing constant. We use Snorkel~\cite{ratner2017snorkel} to learn $\theta$ by minimizing the negative log marginal likelihood given the observed $\Lambda^j$. Finally, as introduced in Ratner et al. \cite{ratner2016data}, the end classifier $f$ is trained with a noise aware loss function that uses probabilistic labels $\hat{Y}^j = p_{\theta}( Y^j| \Lambda^j  )$.

%We next describe how we define heuristics on the time series to detect PVC beats. 
%\subsection

\noindent\paragraph{From Domain Knowledge to an Automated Arrhythmia Detector}
First, we minimally pre-processed the raw ECG signals by removing baseline wandering using a forward/backward, fourth-order high-pass Butterworth filter \cite{lenis2017comparison}. To segment ECG ($x^j$) into individual beats ($x^j_b$), we followed a simple segmentation procedure, where we considered the time segment between two alternate QRS-complexes to be a heartbeat. 

\textbf{\begin{figure}[!htbp]
    \centering
    \includegraphics[width = .9\textwidth]{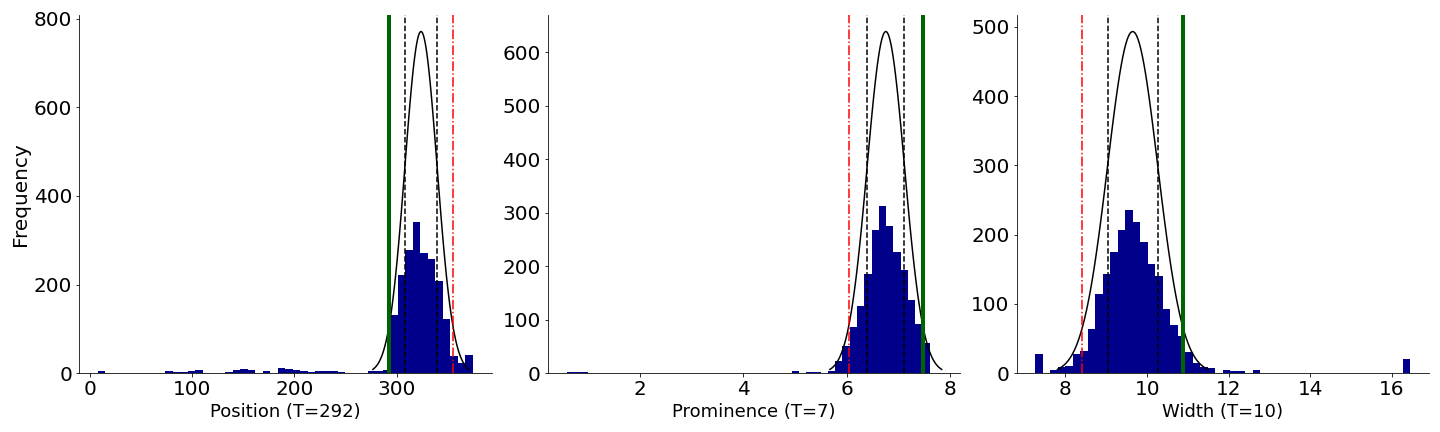}
    \caption{Distribution of the location, height (topological prominence), and width of QRS-complexes of one patient. To turn a loosely defined heuristic such as ``\textit{R-wave appears earlier than usual}'' into an LF, we must characterize the "usual" location of the R-wave. To accomplish this, we we fit a robust Gaussian distribution to model the variance of R-wave locations, and assume any beat located two standard deviations earlier (solid green line at $\text{T}_{\texttt{Early R-wave}} = 292 \ ms$),  than the estimated mean, to be a PVC beat. Since these attributes vary widely from patient-to-patient, we automatically compute these thresholds separately for each patient and heuristic.}
    \label{fig:Tuning}
\end{figure}}

We had to determine the precise locations of the QRS-complexes and T-waves. As in most prior work, we used the approximate locations of the R-wave available for each ECG record in the database, along with Scipy's peak finding algorithm \cite{2020SciPy-NMeth} to find the exact locations of the R and T waves. Further, we used the RANSAC algorithm~\cite{fischler1981random} to fit a robust linear regression line to each ECG record, to determine its baseline (horizontal green lines in Fig.~\ref{fig:PVCandNormalbeat}). The baselines were used to accurately characterize the height and depth of the R and T-waves\footnote[3]{The topological prominence measure returned by Scipy's peak finding algorithm was imprecise.} (blue and red vertical lines in Fig.~\ref{fig:PVCandNormalbeat}). 
% \footnote{\url{https://docs.scipy.org/doc/scipy/reference/generated/scipy.signal.find_peaks.html}}\todo[inline]{better make it a reference not a footnote} 

\begin{figure*}[!hbt]
    \centering
  \begin{tabular}{cc}
    \includegraphics[width=0.47\textwidth]{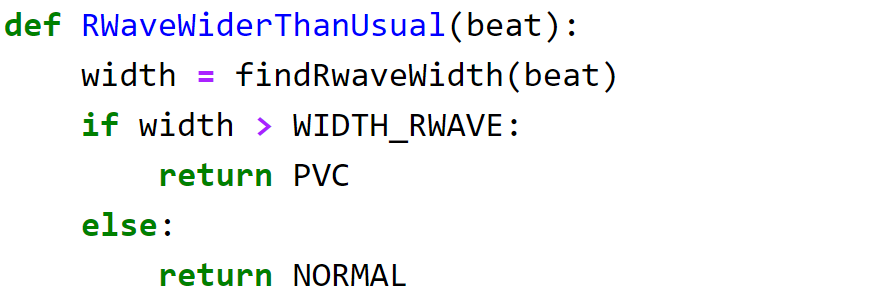}
    &\includegraphics[width=0.47\textwidth]{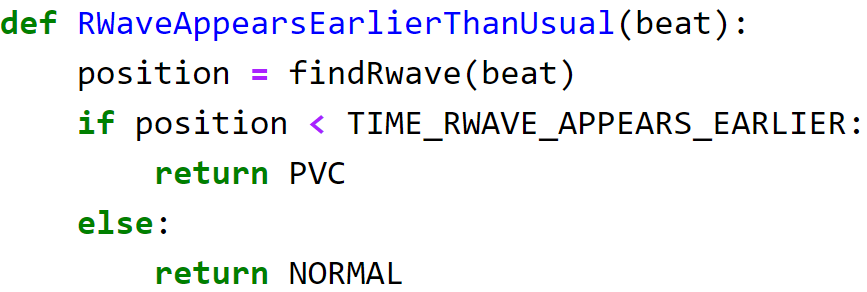}\\
  \end{tabular}
    \caption{Example Python code for $\text{LF}_{\texttt{Wide R-wave}}$ and $\text{LF}_{\texttt{Early R-wave}}$. The \texttt{findRwaveWidth()} and \texttt{findRwave()} sub-routines return the precise width and positions of the R-wave in a beat, while the variables \texttt{WIDTH\_RWAVE} and \texttt{TIME\_RWAVE\_APPEARS\_EARLIER} reflect the thresholds $\text{T}_{\texttt{Wide R-wave}}$ and $\text{T}_{\texttt{Early R-wave}}$.}
    \label{fig:labelingfunctions}
\end{figure*}

Next, we defined $6$ simple LFs based on the domain knowledge
%- described at the beginning of the methodology section - 
to assign probabilistic labels (\texttt{PVC}, \texttt{OTHER} or \texttt{ABSTAIN}) to each beat.  Fig.~\ref{fig:labelingfunctions} provides example pseudocodes for two of the LFs that were defined. To express the loosely-defined domain knowledge we described previously as LFs, we have to automatically assign thresholds to them. For instance, one heuristic to identify a PVC beat is to check whether its ``\textit{R-wave appears earlier than usual}''. To turn this heuristic into a LF ($\text{LF}_{\texttt{Early R-wave}}$), one has to determine the ``usual'' position of the R-wave. For this, we used the Minimum Covariance Determinant algorithm~\cite{rousseeuw1999fast} to find the covariance of the most-normal subset of the frequency histogram. We then set the threshold to the value $2$ standard deviations away from the estimated mean in the direction of interest. For example, for a particular subject (Fig.~\ref{fig:Tuning}) $\text{LF}_{\texttt{Early R-wave}}$ returns \texttt{PVC} for any beat with the R-wave appearing earlier than $\text{T}_{\texttt{Early R-wave}} = 238$ ms (vertical green line). To account for inter-patient variability, we automatically compute these subjective thresholds for each heuristic and every patient separately. Note that some of our heuristics did not require estimating any subject-specific parameters.

\paragraph{The End-Model Classifier}
With these heuristics, we use the label model in Eq.~(\ref{eq:labelmodel}) to obtain probabilistic labels for heartbeats of all training patients in the MIT-BIH Arrhythmia Database. %We trained separate models for different patients since their LFs varied quite drastically due to differing thresholds. 
We use these probabilistic labels and the segmented beats to train a noise-aware ResNet classifier, in which we weigh each sample according to the maximum probability that it belongs to either class. 
%, such that beats which had more or less the same probability of belonging to either class were weighed less than those which DP believed were definitely \texttt{PVC} or \texttt{OTHER} beats
Recent studies have shown that ResNet not only performs on par with most of the state-of-the-art time series classification models~\cite{fawaz2019deep}, but also works well for automatic arrhythmia detection~\cite{ebrahimi2020review}. 

\section*{Experiments and Results}
\label{sec:experiments}

\paragraph{Data}
%\subsection{}
%\label{sec:MITBIH}
The Massachusetts Institute of Technology – Beth Israel Hospital (MIT-BIH) Arrhythmia Database~\cite{moody2001impact} is one of the most commonly used datasets to evaluate automated Arrhythmia detection models. It contains $48$ half-hour excerpts of two-lead ECG recordings from $47$ subjects. In most records, the first channel is the Modified Limb lead II (MLII), obtained by placing electrodes on the chest. We only used the first channel to detect PVC events, since the QRS-complex is more prominent in MLII. The second channel is usually lead V1, but may also be V2, V4 or V5 depending on the subject. We refer the interested reader to Moody et al.~\cite{moody2001impact} for more details on how the database was curated and originally annotated. 

\noindent\paragraph{Experimental Setup}
Our experimental setup follows the evaluation protocol of arrhythmia classification models stipulated by the American Association of Medical Instrumentation (AAMI) as described in~\cite{aami2008r}. The AAMI standard, however, does not specify which heartbeats or patients should be used for training classification models, and which for evaluating them~\cite{luz2016ecg}. Hence, we used the inter-patient heartbeat division protocol proposed by De Chazal et al.~\cite{de2004automatic} to partition the MIT-BIH Arrhythmia Database into subsets DS1 and DS2 to make model evaluation realistic. Furthermore, in the MIT-BIH Database, PVCs only account for $8\%$ of the $100,000$ beats, thus to prevent issues stemming from high class imbalance, we randomly oversampled \texttt{PVC} beats in DS1 before using it to  train the ResNet classifier. The architecture of our ResNet models is the same as in Fawaz et al.~\cite{fawaz2019deep}. To tune the learning rate, batch size, and number of feature maps hyper-parameters, we split the training data into train and validation subsets in the $70/30$ proportion. All the models were trained for $25$ epochs. In the next subsection, we report the results of the ResNet models which had the best true positive rate (TPR) at low false positive rate (FPR) on the validation data.

We also compare the performance of our weakly supervised model with an active learning (AL) alternative. For AL, we used ResNet to iteratively identify data points for manual labeling using uncertainty sampling~\cite{settles2012active}. The model initially had access to a randomly sampled balanced seed set of 100 labeled data points. In each AL iteration, we retrained ResNet using the training data extended with 100 newly labeled data points. We continued this process until the training set consisted of 4,000 points. AL hyper-parameters (the query size and size of the seed set) are similar to Pasolli and Melgani's setup~\cite{pasolli2010active}. Table~\ref{tab:results} reports the performance of the final ResNet model trained on 4000 data points incrementally labeled using AL, averaged over 10 random initializations of the seed set.

\begin{table*}[!htbp]
    \centering
    \resizebox{\textwidth}{!}{
    \begin{tabular}{c|c c c c c c c c c}
         \Xhline{1pt}
         \textbf{Model} & $\textbf{TPR}$ & $\textbf{TNR}$ & $\textbf{PPV}$ & $\textbf{FPR}$ & $\textbf{Acc}$ & $\textbf{FPR}_{50\% \ \text{TPR}}$ & $\textbf{FNR}_{50\% \ \text{TNR}}$ & $\textbf{TPR}_{1\% \ \text{FPR}}$ & $\textbf{TNR}_{1\% \ \text{FNR}}$ \\ \hline
        %  \textbf{Maj.\ labels} & 0.453 & 0.979 & 0.599 & 0.020 & 90.28 & 0.020 & 0.019 & 0.439 & 0.448 \\
         \textbf{Fully sup.} & 0.884 & 0.970 & 0.664 & 0.030 & 96.25 & 0.005 & 0.028 & \textbf{0.793} & 0.266 \\
         \textbf{Pr.\ labels} & 0.645 & 0.960 & 0.523 & 0.039 & 85.84 & 0.019 & 0.140 & 0.165 & 0.252 \\
         \textbf{Active learn.} & 0.514 & \textbf{0.993} & \textbf{0.821} & \textbf{0.007} & 94.15 & 0.020 & 0.021 & 0.604 & 0.405 \\
          \textbf{Weak sup.} & \textbf{0.892} & 0.965 & 0.629 & 0.036 & \textbf{97.25} & \textbf{0.004} & \textbf{0.013} & 0.707 & \textbf{0.466} \\ \Xhline{1pt}
    \end{tabular}}
    \caption{Results on held-out test set. Weakly supervised ResNet performs on par with the fully supervised model and outperforms ResNet trained using active learning. $\textbf{FPR}_{50\% \ \text{TPR}}$ and $\textbf{FNR}_{50\% \ \text{TNR}}$ represent the FPR and FNR at $50\%$ TPR and TNR, respectively. Similarly, $\textbf{TPR}_{1\% \ \text{FPR}}$ and $\textbf{TNR}_{1\% \ \text{FNR}}$ represent the TPR and TNR at $1\%$ FPR and FNR, respectively. The reported AL results are averaged over 10 independent initializations of the random seed set. All measures are computed with \texttt{PVC} as the positive class.} 
    \label{tab:results}
\end{table*}

\noindent\paragraph{Results}
% In this section, we evaluate the efficacy of our weakly supervised arrhythmia detection model. 
We trained ResNet models on DS1 as a training set, using either probabilistic labels or the full ground truth, and evaluated them on the held-out set DS2. The results, summarized in Tab.~\ref{tab:results}, reveal that the end classifier trained using weak supervision is competitive with to the model trained on the full ground truth data. Moreover, our weakly supervised model also outperformed the ResNet trained using 4,000 data points obtained via active learning. 

\begin{figure*}[!hbt]
    \centering
  \begin{tabular}{cccc}
    \includegraphics[width=0.23\textwidth]{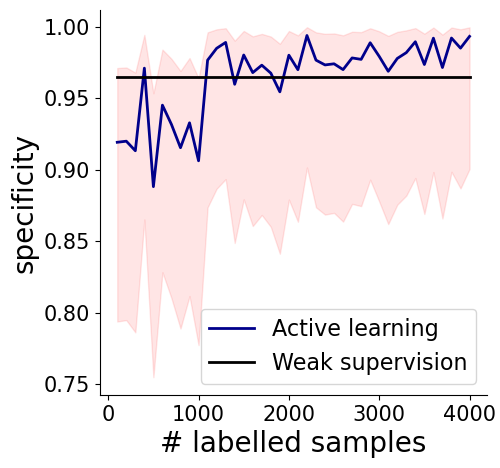} 
    &\includegraphics[width=0.23\textwidth]{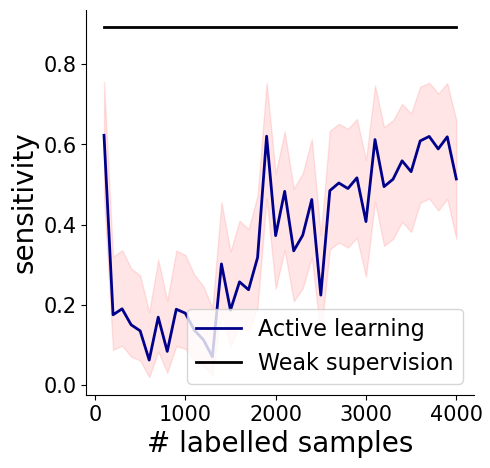} 
    &\includegraphics[width=0.23\textwidth]{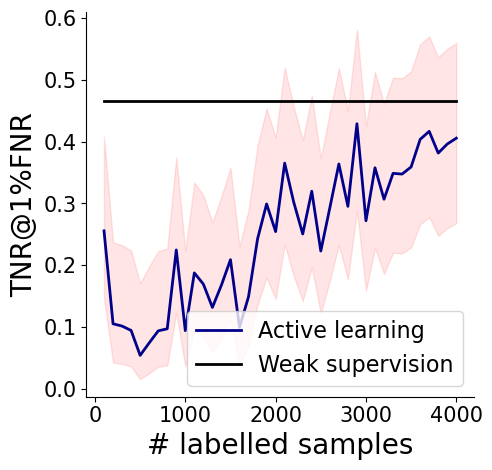}
    &\includegraphics[width=0.23\textwidth]{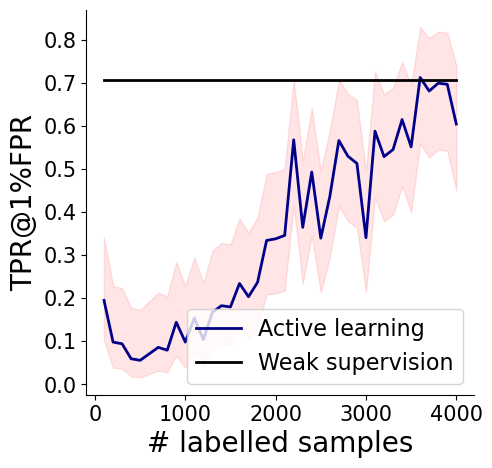}\\
    % (i)&(ii)&(iii)&(iv) \\
  \end{tabular}
    \caption{Active learning results. Our weakly supervised model either exceeds or matches the performance of its AL counterpart. The shaded red regions correspond to the 95\% Wilson's score intervals.}
    \label{fig:Activelearningresults}
\end{figure*}

Let us review the key insights stemming from these results. First, the thresholds for the labeling functions that were automatically determined by our proposed auto-thresholding algorithm varied quite drastically across subjects. For instance, the threshold on the position of the R-wave, $\text{T}_{\texttt{Early R-wave}}$ had a mean of $230 \ ms$ and a standard deviation of $77.14 \ ms$. 
This simple personalization of the LF parameters turned out to be the key to good generalization properties of the end-model;  it failed to perform well when these parameters were fixed to reasonable global settings. 
%generalize well on other subjects if thresholds are fixed. However, using our automatic tuning procedure and only $6$ simple heuristics, we were able to obtain quite accurate probabilistic labels with high specificity. 
The auto-thresholding algorithm is a practically important contribution of our work, at it allows our methods to scale across diverse cohorts of subjects while mitigating potentially excessive manual effort in tuning LFs to specific patients. However, unsurprisingly, even with auto-tuning, our LFs and the estimated probabilistic labels (denoted {``Pr.\ labels"} in Tab.~\ref{tab:results}) were not perfect. In fact, we observed high variability in the performance of Pr.\ labels across different subjects, when compared to ground truth. For example, while they had almost perfect sensitivity for Subject 228 ({TPR} = 0.994), they performed extremely poorly for Subject 214 ({TPR} = 0). Overall, Pr.\ labels had low TPR and high TNR on their own on the training set and held-out test set, which is understandable given the prior class imbalance. 

%  as well as not escalating the risk of overfitting of the downstream models

\begin{figure*}[!hbt]
    \centering
  \begin{tabular}{ccc}
    \includegraphics[width=0.3\textwidth]{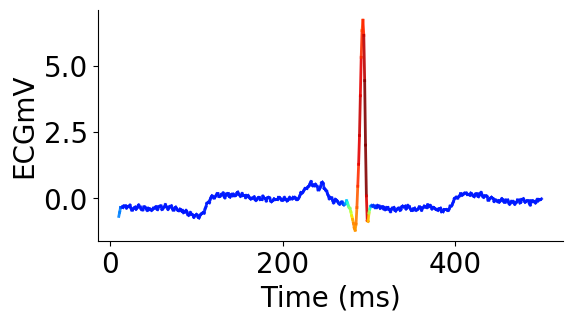} 
    &\includegraphics[width=0.3\textwidth]{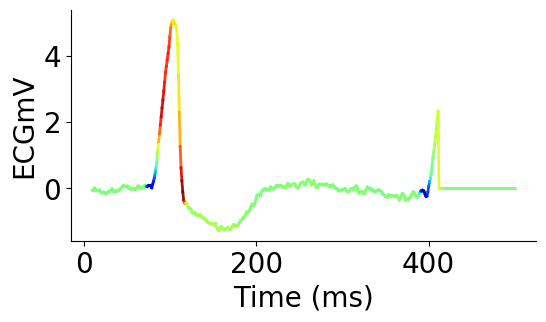} 
    &\includegraphics[width=0.3\textwidth]{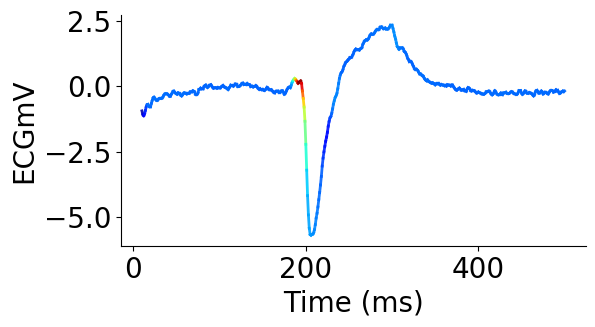} \\
    (i)&(ii)&(iii) \\
    \includegraphics[width=0.3\textwidth]{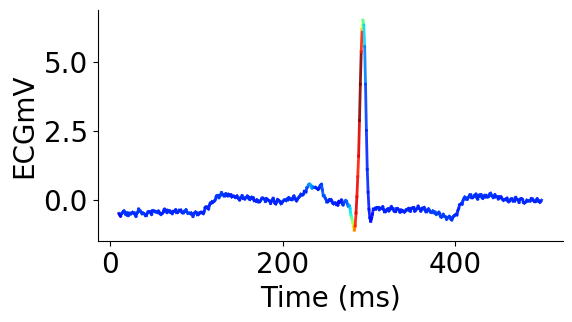} &\includegraphics[width=0.3\textwidth]{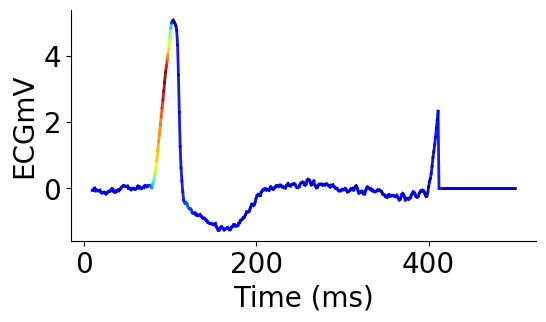} 
    &\includegraphics[width=0.3\textwidth]{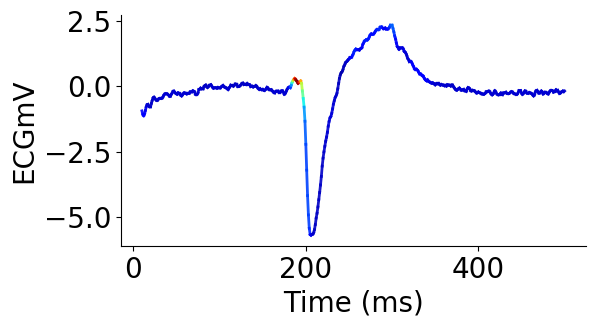}\\
    (iv)&(v)&(vi) \\
%    \includegraphics[width=0.3\textwidth]{Normal beat scatter.png}
%    &\includegraphics[width=0.3\textwidth]{PVC Scatter 17109.png}
%    &\includegraphics[width=0.3\textwidth]{PVC Scatter 17468.png} \\
%    (vii)&(viii)&(iv) \\
  \end{tabular}
    \caption{Class Activation Maps for weakly [(i) - (iii)] and fully supervised [(iv) - (vi)] ResNet reveal that the models discriminate between \texttt{PVC} and \texttt{OTHER} beats primarily based on the morphology of the QRS-complex. 
    %Scatter plots [(vii) - (ix)] illustrate the relationship between class activation intensities of the weakly and fully supervised models for the same beats. 
    The models appear to have learned to focus on similar regularities. Graphs (i) and (iv) represent an example \texttt{OTHER} beat, while the others show two examples of \texttt{PVC} beats.}
    \label{fig:PVC&NormalCAM}
\end{figure*}

Tab.~\ref{tab:results} summarizes performance metrics on the held-out test set (True Positive Rate (TPR), True Negative Rate (TNR), Positive Predictive Value (PPV), False Positive Rate (FPR), and Accuracy (Acc)) measured at the 50-50 class likelihood threshold, chosen for consistency with prior literature on the PVC prediction task. The weakly supervised ResNet ({Weak sup.}) significantly improves sensitivity to the {PVC} class compared to just using the LF labels directly ({Pr.\ labels}) to directly predict the test data. This illustrates that WS ResNet is able to generalize effectively beyond the hypothesis learned by the noisy weak LFs. Our end model performs on par with the fully supervised ResNet ({Fully sup.}) trained on the same data but using all the available pointillistic labels in the MIT-BIH Database. 

Tab.~\ref{tab:results} also compares performance of the four models under consideration at operational settings of pragmatic interest in clinical practice, that is at very low error rates. We report the ability to confidently identify positive cases at FPR of 1\%, and the ability to confidently identify negatives at FNR of 1\%. We complement these results with the error rates observed at 50\% probability of detection of both negative and positive cases.
The results show very little operational utility potential from applying the inferred probabilistic labels directly. 
However, our weakly supervised ResNet model trained on those inferred labels is highly competitive to the equivalently structured ResNet trained on the abundant supply of manually annotated data. 
Weakly supervised ResNet appears particularly strong at identifying negative cases, while its positive recall performance is close to the ground-truth based equivalent. 

Next, we compare performance of the weakly supervised ResNet versus ResNet trained using active learning (``Active learn."\ in Tab.~\ref{tab:results}), and it looks better on all performance metrics barring TNR, PPV and FPR. Graphs in Fig.~\ref{fig:Activelearningresults} show that in the range of up to 4,000 pointillistically labeled training data points, weakly supervised models either outperforms or matches its active learning counterpart, but at a drastically lower requirement of human effort.
%It is likely that the class imbalance is one of the primary reasons behind the comparatively weaker performance of AL. Several AL studies~\cite{zhu2007active, japkowicz2002class} have shown that training sets where the target minority class (i.e. \texttt{PVC}) is significantly under-represented in comparison to the non-target majority class, yield end model classifiers with lower accuracy and area under receiver operating characteristic (AUC) curve. On the other hand, we found our weakly supervised model to be more robust to class imbalance as it explicitly models the prior probability distribution of all the classes. 

To closely examine what the weakly and fully supervised ResNet models are learning, we plotted Class Activation Maps~\cite{zhou2016learning} of a normal and PVC beats in Fig.~\ref{fig:PVC&NormalCAM}. It is evident that to discriminate between PVCs and other beats, our models are primarily paying attention to the QRS complexes in these examples. 
Moreover, it also appears from these plots that both models not only perform on par, but they also tend to focus on similar signatures of the ECG signals. 
%The scatter plots in  Fig.~\ref{fig:PVC&NormalCAM} illustrate an almost linear relationship between the class activation intensities of WS and FS ResNet for some sample beats. 
This observation suggests at least some equivalence between the model trained on ground-truth annotation and the one trained on labels inferred from a small number of simple heuristics.
These results reassure us that the more expensive process can be effectively replaced by the proposed framework of weak supervision that uses a few labeling functions based on high-level aspects of domain knowledge derived directly from the time series characteristics.

\section*{Discussion and  Conclusion}%\label{sec:discussion}

We demonstrated that weak supervision with domain heuristics defined directly on time series provides a promising avenue for training medical ML models without the need for large, manually annotated datasets. 
To support this claim, we developed an arrhythmia detection model which performs on par with its fully-supervised counterpart, and does not need point-by-point data annotation.
This weakly supervised model has been developed in a fraction of time that would be required to provide a fully labeled training set.

We only needed a handful of heuristics to infer probabilistic labels sufficient to yield a reliable end model.
These simple heuristics reflected basic clinical intuition that can be gleaned from ECG diagnostics tutorials.
We expect that engaging expert clinicians to harvest additional heuristics would allow further improvements. 
We stipulate that the proposed approach does not only save effort and time, but it also aligns the process of knowledge acquisition from domain experts better with human nature,
than its tedious pointillistic data annotation alternative. Further, we show that domain heuristics can be automatically tuned to patient specific characteristics by defining parameter tuning rules. 
In our example, auto-tuning of ECG waveform interpretations accounts for inter-patient variability, while keeping manual labor at its minimum. 
%and thus represents an important contribution of our work. 
%Finally, we quantify the time and money that our weakly supervised workflow saves in comparison to its fully supervised complement.

The ML community has devised several techniques to overcome the limitations of expensive pointillistic labeling such as intelligently choosing the most informative training samples to label \cite{settles2012active}, combining both labeled and unlabeled data \cite{van2020survey} and harnessing the power of crowds \cite{foncubierta2012ground, orting2020survey}. While semi-supervised learning has been successfully applied to improve arrhythmia detection models without patient-specific data~\cite{zhai2020semi}, these methods still rely on a significant proportion of labeled training data to start with. On the other hand, crowdsourcing has shown promise in generating ground truth for e.g.\ medical imaging, but prior research~\cite{foncubierta2012ground} found several limitations such as the lack of trustworthiness, inability of non-expert workers to annotate fine-grained categories and ethical concerns around patient privacy. Active learning, however, has by far been the most commonly utilized technique in settings where annotating large quantities of data {\em en-masse} is prohibitively expensive \cite{wang2014797}. 

Multiple avenues of future work include modelling dependencies between LFs to improve both the efficiency and accuracy of label models, and developing a library of time series primitives to streamline development of LFs for such data. 
We would also like to build interfaces to support interactive discovery of LFs and to rigorously validate resulting end models.
Further, we intend to investigate hybrid approaches that will opportunistically combine weak supervision with pointillistic active learning, 
and conduct user studies with clinicians to better understand the challenges and opportunities for interactive harvesting of domain knowledge.
We also aim to enable detection of other types of abnormalities that can be seen in ECG data, and apply our approach to other types of hemodynamic monitoring waveforms.

Time series data is prevalent in healthcare. However the costs of preparing such data for training and validation of new models, as well as for the maintenance of already developed models,
prohibit the otherwise realizable benefits from widespread adoption of machine learning in clinical decision support.
We believe that approaches similar to the one presented in this paper could help making a decisive push towards
proliferating beneficial uses of machine learning in this important field of its application.

\section*{Acknowledgements}
This work was partially supported by the Defense Advanced Research Projects Agency award FA8750-17-2-0130, and by the Space Technology Research Institutes grant from National Aeronautics and Space Administration’s Space Technology Research Grants Program.

\makeatletter
\makeatother

\bibliographystyle{unsrt}
\bibliography{bibliography}

\end{document}